\documentclass[12pt]{elsarticle}

\RequirePackage[left=0.8in,right=0.8in,top=1in,bottom=1in]{geometry}
\usepackage[utf8x]{inputenc}
\usepackage{setspace}
\usepackage{amsmath}
\usepackage{charter}
\usepackage{helvet}

\usepackage{graphicx}
\usepackage{subfigure}
\usepackage{gnuplottex}
\usepackage{amssymb}
\usepackage{booktabs}
\usepackage{lipsum}

\usepackage{nomencl} 
\makenomenclature
 \usepackage[belowskip=-15pt,aboveskip=0pt]{caption}

\usepackage[compact]{titlesec}

\begin{document} 
\begin{frontmatter}
\title{Analyses and estimation of certain design parameters of micro-grooved heat pipes}
\author{Ramachandran.R\(^{2}\)}
\author[rvt]{S.Anil Lal\corref{cor1}}
\ead{anillal@cet.ac.in}
\address{\(^{1}\)Department of Mechanical Engineering, College of Engineering Trivandrum, Kerala, India}
\address{\(^{2}\)LBS Insti­tute of Tech­nol­ogy for Women,Thiru­vanan­tha­pu­ram}

\begin{abstract}
{\it A numerical analysis of heat conduction through the cover plate of a heat pipe is carried out to  determine  the temperature  of the working substance, average temperature of heating and cooling surfaces,  heat spread in the transmitter, and the heat bypass through the cover plate. Analysis has been extended for the estimation of heat transfer requirements at the outer surface of the condenser under different heat load conditions  using Genetic Algorithm. This paper also presents the estimation of an average heat transfer coefficient for the boiling and condensation of the working substance inside the microgrooves corresponding to a known temperature of the heat source. The equation of motion of the working fluid in the meniscus of an equilateral triangular groove has been presented from which a new term called the minimum surface tension required for avoiding the dry out condition is defined.  Quantitative results showing the effect of  thickness of cover plate, heat load, angle of inclination and viscosity of the working fluid on the  different aspects of the heat transfer,  minimum surface tension required to avoid dry out, velocity distribution of the liquid, and radius of liquid meniscus inside the micro-grooves have been  presented and discussed.}
\end{abstract}
\end{frontmatter}
\begin{thenomenclature} 
 \nomgroup{A}
  \item [{$A_{l}$}]	\begingroup Area of liquid ($m^2$).				\nomeqref {1} 	\nompageref{1}
  \item [{$f$}]		\begingroup Friction factor.					\nomeqref {1} 	\nompageref{1}
  \item [{$g$}]		\begingroup Acceleration due to gravity ($m/s^2$).		\nomeqref {1} 	\nompageref{1}
  \item [{$h$}]		\begingroup Heat Transfer Coefficient ($W/m^2 K$).		\nomeqref {1} 	\nompageref{1}
  \item [{$h_{eq}$}]	\begingroup Equivalent Film heat transfer coefficient($W/m^2 K$).\nomeqref {1}	\nompageref{1}
  \item [{$h_{fg}$}]	\begingroup Latent heat of vapourisation ($W/m^2 K$).		\nomeqref {1}	\nompageref{1}
  \item [{$k$}]		\begingroup Thermal conductivity of cover plate ($W/mK$).	\nomeqref {1}	\nompageref{1}
  \item [{$L_{hp}$}]	\begingroup Length of the Heat Pipe ($m$).			\nomeqref {1}	\nompageref{1}
  \item [{$L_{w}$}]	\begingroup Total wetted length ($m$).				\nomeqref {1}	\nompageref{1}
  \item [{$Nu$}]	\begingroup Convective Nusselt number.				\nomeqref {1}	\nompageref{1}
  \item [{$q$}]		\begingroup Heat flux into the working fluid ($W/m^2$).		\nomeqref {1}	\nompageref{1}
  \item [{$R$}]		\begingroup Radius of the liquid meniscus ($m$).		\nomeqref {1}	\nompageref{1}
  \item [{$T$}]		\begingroup Temperature ($K$).					\nomeqref {1}	\nompageref{1}
  \item [{$T_{evp}$}]	\begingroup Evaporator temperature ($K$).			\nomeqref {1}	\nompageref{1}
  \item [{$T_i$}]	\begingroup Interface Temperature($K$).				\nomeqref {1}	\nompageref{1}
  \item [{$T_{\infty}$}]\begingroup Ambient temperature($K$).				\nomeqref {1}	\nompageref{1}
  \item [{$t$}]		\begingroup Thickness of cover plate $(mm)$.			\nomeqref {1}	\nompageref{1}
  \item [{$W$}]		\begingroup Wetted Length $(mm)$.				\nomeqref {1}	\nompageref{1}
  \item [{$z$}]		\begingroup Height of the liquid meniscus ($m$).		\nomeqref {1}	\nompageref{1}
  \item [{$\sigma$}]	\begingroup Surface tension ($N/m$).				\nomeqref {1}	\nompageref{1}
  \item [{$\sigma_{r}$}]\begingroup Surface tension required to avoid dry out ($N/m$).	\nomeqref {1}	\nompageref{1}
  \item [{$\rho$}]	\begingroup Density of working substance ($kg/m^{3}$).		\nomeqref {1}	\nompageref{1}
  \item [{$\beta$}]	\begingroup Inclination of the heat pipe with horizonal ($deg$).\nomeqref {1}	\nompageref{1}
  \item [{$\nu$}]	\begingroup Kinematic viscosity ($m^{2}/s$).			\nomeqref {1}	\nompageref{1}
\end{thenomenclature}

%
%
\doublespacing

\section{INTRODUCTION}
\begin{figure}[h]
\begin{center}
\begin{center}\includegraphics[scale=0.8]{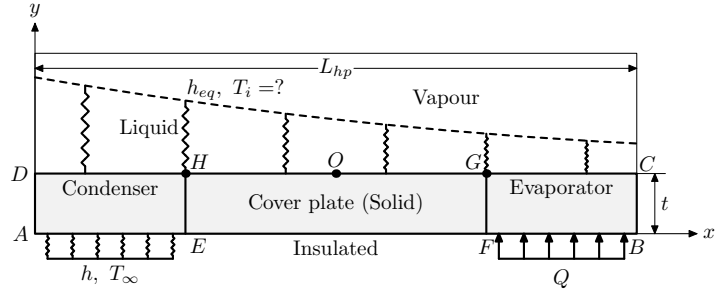}\end{center}
\end{center}
\caption{Schematic diagram showing the heat transfer in a heat pipe.}
\label{scheme}
\end{figure}

\begin{figure}[h]
\begin{center}
\begin{center}\includegraphics[scale=0.8]{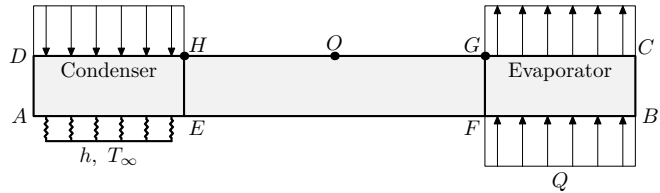}\end{center}
\end{center}
\caption{A uniform heat flux at the evaporator and condenser  and zero heat flux through the adiabatic transmitter.  This is closely applicable for small thickness of cover plate \(ABCD\) and high convective film heat transfer coefficients.} 
\label{unihf}
\end{figure}

\begin{figure}[h]
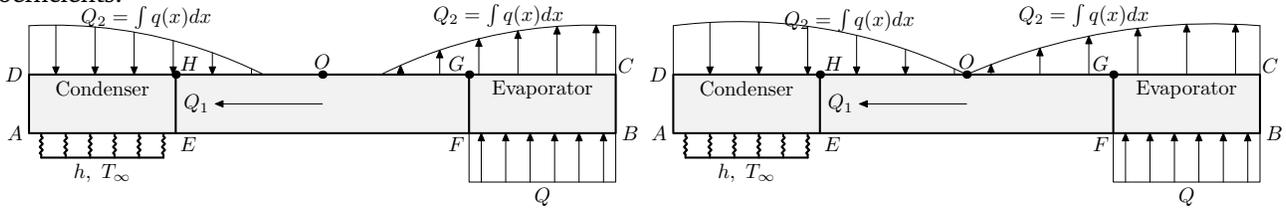

\begin{center}
\includegraphics[scale=0.78]{c3.mps}
\includegraphics[scale=0.78]{c5.mps}
\end{center}
\caption{Typical distribution of heat flux over the top surface \(CD\) of the cover plate, for cases with axial heat conduction and heat spread. The net heat flow across \(CD\) is zero and \(Q=Q_1+Q_2\).} \label{dis}
\end{figure}

Heat pipe is a device used for absorbing heat from a heat source, transmitting it through a required distance and rejecting it to a sink. It essentially consists of a working fluid undergoing a phase change and the  portions of the heat pipe in which different processes take place are called evaporator, transmitter and condenser. Present work is related to the analysis of the heat transfer process through the solid parts and working substance of heat pipes aimed at finding the favorable parameters for improving the design.  A schematic diagram showing the heat transfer process through the cover plate of a typical heat pipe is given in Figure~\ref{scheme}. The temperature  of the outer surface of the evaporator, where the heat source is placed is the important parameter to be optimized in thermal management of equipments/devices using heat pipes. Its value is governed by the heat transfer coefficient inside the channel, thermal conductivity of the materials of the cover plate and wick, heat transfer coefficient outside the condenser, thickness of the cover plate and temperature at which the working substance condenses and evaporates denoted as \(T_i\) in figure \ref{scheme}.  By considering a heat conduction analysis, computation of a uniform value for \(T_i\) of the working substance temperature along \(DC\) under steady state condition (ie, no heat accumulation in the working substance) is possible. In other words, by taking the net heat transfer across \(DC\) as zero, it is possible to find out a uniform value for \(T_i\) and hence a non-uniform distribution of heat flux along \(x-\)direction denoted as \(q(x)\) in Fig~\ref{dis}, corresponding to  a finite value of thickness of the cover plate \(ABCD\).  Do et al. \cite{do2008mathematical} have noted that several prior investigations have assumed a uniform heat flux into the evaporator and condenser and no heat transfer in the transmitter  as illustrated in Figure \ref{unihf}. The transmitter of a heat pipe is normally insulated from outside and is commonly referred to as adiabatic transmitter section. Vadakkan et al. \cite{vadakkan2003transient, vadakkan2004transport} have demonstrated that there occurs heat transfer to the working substance at the axial locations of adiabatic transmitter as well, due to spread of the heat flux in the axial direction. The distribution of heat flux \(q(x)\) into the fluid from the inner surface of the cover plate depends on the temperature distribution of the cover plate and the constant liquid-vapor interface temperature (\(T_i\))of the working substance.  It is noted in ref. \cite{anand2002experimental} that the flow rate, velocity and thickness of the liquid film of the working substance inside  micro channels of  heat pipes is controlled by the distribution heat flux, \(q(x)\) into the micro channels. The heat flux through the cover plate may have axial and transverse (across the thickness) components due to axial conduction and heat transfer to the working substance respectively. A brief review of the literature dealing with computation of the liquid-vapor interface temperature of the working substance, the heat flux distribution into the working substance and axial heat conduction through the plate are discussed below. 

Aghvami and Faghri \cite{aghvami2011analysis} have presented an analysis of heat conduction through the cover plate of a flat heat pipe by solving two-dimensional heat conduction equation. They specified a constant  temperature for the liquid vapor interface and used a linear interpolation to get the inner wall temperature. This is equivalent to treating the wick and liquid within the wick as a thermal resistance. The constant liquid-vapor interface temperature is found out by solving separate equations in liquid and vapor regions. Hence the heat transfer to the liquid and axial heat conduction depend solely on the value of the temperature prescribed at the inner surface of the cover plate.  Sobhan et al. \cite{sobhan2000investigations} have employed a transient computational methodology using ambient temperature as the initial temperature and presented the results for steady state conditions as well. In this study a constant liquid-vapor interface temperature computed using Clausius-Clapeyron equation is taken as the boundary condition. Vadakkan et al. \cite{vadakkan2004transport} have also used a transient approach in which the wick-vapor interface temperature is computed from  energy balance at the interface and liquid pressure at the interface is computed  using  Clausius-Clapeyron equation. \\ 
In the case of micro heat pipes, the  cover plate is very thin and the assumption of uniform evaporation and condensation in the respective portions and no phase change in the adiabatic region as illustrated in Figure \ref{unihf} is generally followed. Sobhan et al. \cite{sobhan2000investigations} have investigated the transient and steady state performance of a micro heat pipe with triangular channels by considering the heat exchange across the evaporator and condenser  as  constant values. This heat pipe does not have a adiabatic transmitter. The results show that distribution of velocity in the channel has a sharp change of slope. One of the reasons for this sharp variation of velocity in the channel is the  use of uniform heat flux  at the condenser and the evaporator. An analytic model was developed by Suman et al. \cite{suman2005model} to find expressions for the critical heat input, dry out length and available capillary head for the fluid flow. The heat pipe studied has an adiabatic transmitter in between the evaporator and condenser. This study also used a discontinuous variation of heat flux along the length with uniform values at the evaporator and condenser and zero heat flux at the adiabatic transmitter. Launay et al. \cite{launay2004hydrodynamic} have studied hydrodynamic and thermal  behavior of a water-filled micro heat pipe. They considered a uniform distribution of heat flux in the evaporator, neglected the axial conduction through the wall and applied a constant temperature along the inner wall of the heat pipe. The equilibrium temperature of the liquid-vapor interface is taken as a constant value more than the saturation temperature of the working substance.  Longtin et al. \cite{longtin1994one} presented a one dimensional model of a micro heat pipe operating under steady state conditions, for predicting the fluid thermal behavior including maximum heat transfer capability, effect of pipe length, width, working fluid and optimum operating conditions. They assumed a constant vapor temperature, a uniform evaporator heat flux and also assumed that adiabatic transmitter experiences no heat transfer. The results show that maximum heat transfer capability of a micro heat pipe varies as inverse of its length and cube of its hydraulic diameter. \\ 
Sartre et al. \cite{sartre2000effect} presented a three dimensional steady state model for predicting the effect of inter-facial phenomenon on evaporative heat transfer in micro heat pipes. They applied a uniform heat flux on the upper surface of the evaporator plate and its lower surface is cooled by natural convection. They proved that a  part of the total heat input in the evaporator passes through the cover plate.  Hung and Tio \cite{hung2010analysis} have investigated the effect of axial conduction through the solid walls of a micro heat pipe using an analytic solution of a one dimensional steady state model. In this study, the temperature distribution of the solid is found as a function of a constant liquid temperature by considering a convective resistance  corresponding to \(Nu=2.68\) for the convective films at the evaporator and condenser. Then they proved that the liquid temperature is equal to the average solid wall temperature by equating the heat added and removed at the evaporator and condenser. They also observed a significant axial heat conduction at the walls for different materials such as  Copper, Nickel and Monel. \\ 

The foregoing literature shows the importance of the analysis of heat conduction through the cover plate for the  prediction of important parameters related to the design and the operation of heat pipes. Also it is found from literature that, the interface temperature of the evaporating liquid has been determined either by thermodynamic considerations \cite{vadakkan2004transport},\cite{sobhan2000investigations} or by using heat transfer considerations as in \cite{aghvami2011analysis}. It is noted that a uniform temperature computed from analysis of heat transfer through the cover plate will be accurately provided, evaporation of the working substance is thermodynamically  possible under the temperature and other conditions applied. Therefore, the present work is concerned with development of a numerical scheme for the determination of a constant liquid-vapor interface temperature (\(T_i\)) of the working substance by analyzing the heat transfer through the cover plate. Further, the analysis provides the temperature distribution in the cover plate and data for the determination of distribution of heat flux into the working substance, the amount of heat bypassing the working substance, if any, etc. The heat flux distribution is an important input to determine the temperature of working substance inside the channel.  To enable the design of the heat transfer process to be implemented over the cooling side of the condenser, an inverse heat conduction problem to estimate the heat transfer coefficient over the outer surface of the condenser  to limit the average heater surface temperature (denoted as \(T_{evap}\)), below a prescribed value is  solved.  This paper describes the numerical method for the solution of the liquid-vapor interface temperature, the methodology involving  Genetic Algorithm (GA) for  solving the inverse heat conduction, equation of motion of liquid in the meniscus of a micro-sized triangular grooved channel and the methodology of solution for liquid velocity and meniscus radius. The effect of varying the thickness of the cover plate on the heat transfer coefficient outside the condenser has been studied. From the analysis of flow through the working substance through  micro-grooves, the minimum value of surface tension required (denoted as \(\sigma_r\)) for making the flow under a given set of conditions has been determined. This is another data useful for the selection of right working substance for the heat transfer.
\section{MODELING OF HEAT CONDUCTION}
In Figure \ref{scheme}, the rectangle \(ABCD\) represents the metallic cover plate of a heat pipe that  separates the working substance and the heat exchanging surfaces. The edge \(FB\) represents the evaporator of the heat pipe through which a uniform heat flux enters the cover plate. The condenser is represented by the cooling edge \(AE\), where a fluid at temperature \(T_{\infty}\)  with a convective   heat transfer coefficient \(h\) removes heat. The edge \(EF\) is insulated and the edge \(DC\) is exchanging heat to the working substance. It is assumed that the liquid vapor interface remains steady so that the net heat transfer across \(CD\) is zero for a given constant heat flux through \(FB\). In this manner  the edge \(DC\) is exchanging heat to a constant temperature reservoir which acts  simultaneously as a heat sink and heat source. An equivalent heat transfer coefficient denoted as \(h_{eq}\) for the exchange of heat from the top edge \(DC\) to the constant temperature liquid vapor interface is considered. It has been 
reported by Ma and Peterson \cite{ma1997temperature} that for triangular grooved channels, the heat transfer coefficient for heat exchange with the boiling liquid over \(CD\) could  vary typically in the range \(3.7-6.7~W/cm^2K\). For flat heat pipes an equivalent heat transfer coefficient should be computed  based on the thermal conductivity of the materials of liquid  and wick and the liquid film  heat transfer coefficient. The heat transfer problem consists of solving the constant temperature of the liquid-vapor interface, \(T_{i}\) and determination of the distribution of heat flux, \(q(x)\) to the liquid through \(CD\).   \\ 
From the earlier section it is noted that  most of the previous investigations on flat heat pipes and micro heat pipes have assumed the resistances of the metallic plate and convective film resistances as zero and  consequently, a heat flux distribution similar to the one  shown in Figure \ref{unihf} is used. Determination of  \(q(x)\) for a known value of liquid temperature \(T_{i}\) is a  heat transfer problem which can be  solved directly. But in the present work \(T_i\) is an unknown. Corresponding to phase change of the working substance in evaporation and condensation, the temperature \(T_i\) should have a constant value over the entire length of the cover plate. Another known condition is that the net heat exchange to liquid through the edge \(CD\) is zero, which can be mathematically expressed as \(\int_{C}^{D}q(x)~dx=\int_{C}^{O}q(x)~dx-\int_{D}^{O}q(x)~dx= Q_2-Q_2=0\). Typical variations of heat flux \(q(x)\) along the edge \(CD\) of the cover plate are shown in  Figure \ref{dis}.

The mathematical model of the problem is the Laplace equation of temperature at all the points within the domain subjected to boundary conditions. The governing equation and boundary conditions are as in the following
\begin{eqnarray}
 &&\nabla^2T=0~~~\mbox{at all the points inside the cover plate} \label{g1}\\
 &&k\frac{\partial T}{\partial y}=-h(T-T_{\infty}) ~~\mbox{and}~~\int_{F}^{B}k\frac{\partial T}{\partial y}~dx=-Q  \label{b1} 
 \end{eqnarray}
 at points on AE and on FB respectively
 \begin{eqnarray}
 \mbox{At point on CD,~} Q_2&=&-\int_{O}^{C}k\frac{\partial T}{\partial y}dx=\int_{D}^{O}k\frac{\partial T}{\partial y}dx \\
 \mbox{Such that~}\int_{D}^{C}k\frac{\partial T}{\partial y}dx&=&0  \nonumber
 \end{eqnarray}
 Referring to Figure (\ref{scheme}),along \(D\) to \(C\),
 \begin{eqnarray}
 &&\Rightarrow \int_{D}^{C}k\frac{\partial T}{\partial y}dx=-\int_{D}^{C} h_{eq}(T-T_i)dx = 0\\
 &&\Rightarrow \int_{D}^{C}\left(k\frac{\partial T}{\partial y}+ h_{eq}(T-T_i)\right)dx = 0  \label{b2}
 \end{eqnarray}
Where, \(h_{eq}\) is an equivalent heat exchange coefficient. In the case of micro heat pipes \(h_{eq}\) is the average value of the heat transfer coefficient in boiling and condensation.
\section{NUMERICAL METHODOLOGY FOR HEAT TRANSFER}
In the present work a cell centered finite volume method is applied for the solution of equation (\ref{g1}). Referring to the standard grid system for a general variable as followed by Patankar \cite{patankar1980numerical},  the  discretization equation for the control volume with geometric center at \(P\), with East, North, West and South neighbors denoted by \(E\), \(N\), \(W\) and \(S\) is
\begin{equation}
 A_PT_P=A_ET_E+A_WT_W+A_NT_N+A_ST_S
\end{equation}
Where \(A's\) are the influence coefficient and \(T\) is temperature. The application of the boundary condition in equation (\ref{b1}) for the solution of the governing equation  is straight forward. But application of the boundary condition on the top line \(CD\) as per equation (\ref{b2}), where the liquid-vapor interface with unknown temperature is the North neighbor requires  special attention and the details is described in the following.
\subsection*{Discretization equation for liquid-vapor interface temperature, \(T_i\)}
\begin{figure}[h]
\begin{center}
\includegraphics[scale=0.8]{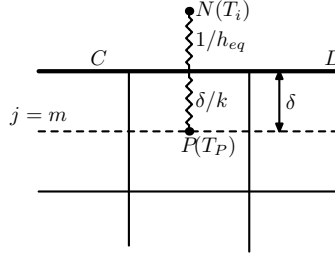}
\end{center}
\caption{Details of the grid point just below the cover plate and the liquid vapour interface temperature,\(T_i\) as its north side neighbor,  considered for the derivation of discretization equation for \(T_i\). } \label{inttemp}
\end{figure}

Figure \ref{inttemp} shows the grid line in the metal plate just below the edge \(CD\) denoted as \(j=m\) with the liquid-vapor interface as the North neighbor . For the control volume shown  as \(P\) on the line \(j=m\), the heat transfer towards the liquid-vapor interface is obtained by integrating equation (\ref{b2}) over the control volume, thereby the rate of heat transfer becomes 
\[\frac{k~h_{eq}\triangle x_P}{k+\delta~h_{eq}}\left(T_P-T_i\right)\]
From this expression for heat transfer, the North influence coefficient of grid points along the line \(j=m\) becomes
\[A_N=\frac{k~h_{eq}\triangle x_P}{k+\delta~h_{eq}}\]
The discretization equation for the computation of \(T_i\) is obtained by taking the total heat transfer across the top edge CD equal to  zero, which can be expressed as
\begin{equation}
\sum_{P}A_N(T_i-T_P) = 0~~\Rightarrow T_{i}\sum A_N=\sum_{P=1}^{n}A_NT_P \label{d1}
\end{equation}
Equation (\ref{d1}) shows that  under steady state condition, the vapor liquid interface temperature is the weighted average of the temperature of grid points below the top line \(CD\). While, the discretization equation of any interior grid point involves the influence of temperature of four neighboring grid points, the discretization equation (\ref{d1}) of \(T_i\) consists of influence of temperature of all the grid points on the grid line below \(CD\). In the present work all the discretization equations including the one for \(T_i\) are solved using conjugate gradient method. 
\section{METHODOLOGY FOR INVERSE HEAT TRANSFER PROBLEM}
By taking the heat transfer coefficient \(h\) at the cooling side of the condenser as a controllable parameter, an inverse problem is considered to   determine the value of \(h\) required for limiting the heater surface temperature on the outside surface of the evaporator to \(T_{evap}\). This value for \(h\) is denoted as \(h_{r,T_{evap}}\). Genetic Algorithm (GA) is applied to estimate the value for  \(h_{r,T_{evap}}\). GA mimic the theory of natural evolution to maximize an objective function. Here the value of \(h_{r,T_{evap}}\) is to be estimated such that the temperature over the surface of the heat source computed from the theoretical model becomes equal to the prescribed minimum value of temperature, denoted as \(T_{evap}\). Hence the objective function is taken as
\begin{equation}
\mbox{Maximize,}~Z=\frac{1}{1+|T_{comp}-T_{evap}|} 
\end{equation}
where \(T_{comp}\) is the average value of temperature of the surface of the heat source computed corresponding to the values of \(h\) assigned to a set of objects in populations that are undergoing changes through generations. The motivation for changes between generations is to  maximize the value of \(Z\). For solving \(T_{comp}\) a subroutine developed based on the numerical method outlined in the previous section has been incorporated as a forward model.  In the present work the value of \(T_{evap}\) is taken as \(350~K\). The computer code in FORTRAN for GA  was adapted from the GA driver created and made available as an open source by Carroll \cite{carroll1996genetic}. The important input parameters of the GA computer code are, size of the population, number of generations, minimum value of the parameter and maximum value of the parameter. As suggested by Carroll \cite{carroll1996genetic}, a uniform cross over and a population size of five is applied. The minimum and maximum values of parameters to be given as input vary with the problem and a narrow range that contains the optimum solution (narrow search space)  gives more accuracy as well as convergence in less number of generations. For the single parameter estimation as that of the present work, a narrow search space can be easily found out by starting from a wide search space and executing the program for a few number of generations (5 generations in the present work). At the end of the first execution of GA in this manner, a suitable narrow range can be identified from the parameters applied on the objects of population and the corresponding computed temperature.   \\ 

Another important parameter influencing the accuracy of heat transfer computation for heat pipes is the equivalent film heat transfer coefficient \((h_{eq})\) over the inside surface of the microgroove, where evaporation and condensation of the working substance is taking place. From literature, it is found that value of this two phase heat transfer coefficient is not reported  with high accuracy  and the uncertainty involved in its experimental determination is high. Cengel \cite{cengel2007heat} reported that the heat transfer coefficient in boiling and condensation ranges from \(2500-100000~W/(m^2K)\). Ma and Peterson \cite{ma1997temperature} reported the same to vary in the range \(37000-67000~W/(m^2K)\). A recent study by Laubscher and Dobson \cite{dobson2013boiling} on the boiling and condensation heat transfer coefficients for a heat pipe heat reports a maximum of \(8000~W/m^2K\) for boiling and \(5000~W/m^2K\) for condensation. The present inverse solution to the heat transfer problem is also used to estimate the value of \(h_{eq}\)  as a function of temperature of the heat source (could be obtained from measurement), convective heat transfer coefficient on the outside surface the condenser and bulk temperature of the cooling fluid, heat load, geometry and material of the heat pipe. It may be noted that all the independent variables are occurring outside the heat pipe, so accurate experimental measurement of these variables is possible.
\section{MODELING OF MOTION OF LIQUID IN THE MENISCUS ON MICRO CHANNEL}
\begin{figure}[h]
\begin{center}
\begin{center}\includegraphics[scale=0.8]{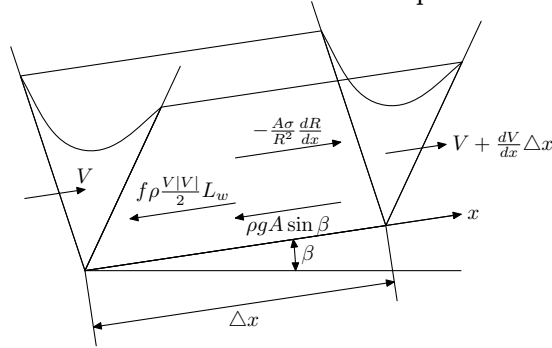}\end{center}
\end{center}
\caption{Differential element of liquid meniscus in a micro grooved channel subjecting to velocity gradient under surface tension, gravity and frictional forces }
\label{mgelem1}
\end{figure}

\begin{figure}[h]
\begin{center}
\begin{center}\includegraphics[scale=0.8]{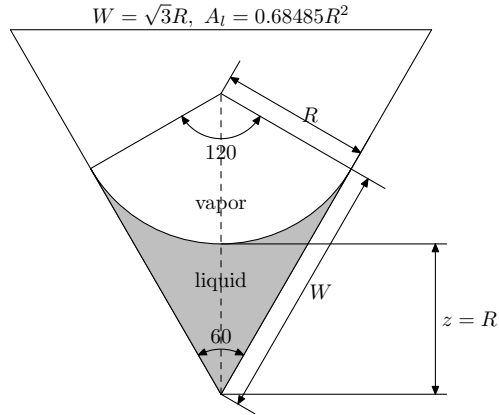}\end{center}
\end{center}
\caption{The geometry of liquid meniscus in a micro grooved channel. Radius of the meniscus \(R(x)\) is same as \(z(x)\). \(A_l\) is the area occupied by liquid depends on \(x\), which is expressed as a function of \(R(x)\). Total wetted length is \(2W=2\sqrt{3}R\).  }
\label{mgelem2}
\end{figure}

Referring to the differential element of a liquid meniscus in a micro-groove in Figure \ref{mgelem1}, the differential equation of the motion of liquid in the meniscus can be written as
\begin{eqnarray}
\rho~A_lV\frac{dV}{dx}+A_l\frac{\sigma}{R^2}\frac{dR}{dx}+f\rho\frac{V|V|}{2}L_{w}+\rho gA_l\sin(\beta)=0 \label{first}
\end{eqnarray}
The forces considered here are due to surface tension, gravity and friction.  For the equilateral triangular channel with the details given in Figure \ref{mgelem2}, the area of liquid flow  can be expressed as,\(A_l=CR^2\), where \(C=0.68485\) and the total wetted length,  \(L_{w}=2\sqrt{3}R\). For this geomtery, \(R\) also represents  the height of liquid meniscus. \\ An equation of  gradient of rate of flow of liquid along the passage, corresponding to a steady liquid-vapour interface during the evaporation and condensation process taking place inside the channel can be written as 
\begin{equation}
 \frac{dQ}{dx}=\frac{d}{dx}(A_lV)=\frac{q(x)}{\rho~h_{fg}} \label{cont1}
\end{equation}
where \(Q\) is the rate of flow and \(Q=A_lV\). The right hand side term of the above equation is known from the calculation of  heat transfer through the cover plate. Therefore,
\begin{equation}
 A_l\frac{dV}{dx}+V\frac{dA_l}{dx}=\frac{dQ}{dx}\Rightarrow \rho A_lV\frac{dV}{dx}=\rho~V\frac{dQ}{dx}-\rho~V^2\frac{dA_l}{dx} \label{three}
\end{equation}
By re-writing the first term of equation (\ref{first}) with the RHS of equation (\ref{three}), and taking \(A_l=CR^2\), it is possible to write
\begin{eqnarray}
\rho~V\frac{dQ}{dx}-2CR\rho~V^2\frac{dR}{dx}+CR^2\frac{\sigma}{R^2}\frac{dR}{dx}+f\rho\frac{V|V|}{2}2\sqrt{3}R+\rho gCR^2\sin(\beta)=0 \label{thick0}
\end{eqnarray}
The friction factor for the flow of liquid through the micro channel is taken as \(\frac{13.33}{Re}\) according to references \cite{sobhan2000investigations},\cite{suman2005model},\cite{suman2005analytical}. Here \(Re\) is computed from the average value of velocity \(\overline{V}\) of liquid and average meniscus radius \(\overline{R}\). The characteristic dimension of flow is taken as the ratio of four times area of flow to wetted length, which for the current geometry in Figure 2 is \(D=0.7907\overline{R}\). Thus the final expression for the friction factor taken for computations is \(f=\frac{16.85\nu}{\overline{R}\overline{V}}\). \\
After rearrangement and substitutions,  equation (\ref{thick0}),  for the gradient  of radius of the liquid meniscus has the following form
\begin{eqnarray}
\frac{dR}{dx}=\frac{\rho/C\left(V\frac{q(x)}{\rho~h_{fg}}+CgR^2\sin(\beta)+\frac{29.19\nu~RV|V|}{\overline{R}\overline{V}}\right)}{(2\rho R V^2-\sigma)}  |\label{thick1}
\end{eqnarray}
Equations (\ref{cont1}) and (\ref{thick1}) are two first order differential equations of \(Q\) and \(R\). After solving \(Q\) and \(R\), the liquid velocity is computed by \(V=Q/A_l=Q/(CR^2)\). The side of the heat pipe, where the condenser starts is taken as \(x=0\) and the initial value problem is solved by a two step numerical integration using Adam-Bashforth method. \\
By noting that,  positive \(x-\)direction is from condenser to evaporator, the sign of \(\frac{dR}{dx}\) as per equation (\ref{thick1}) should be negative and its magnitude must be less than a specified value for avoiding dry out (\(R(x)\) becoming equal to zero at \(x<L_{hp}\)). Further, all the terms of the numerator of this equation are positive. This leads to the condition that \(\sigma-2\rho~RV^2\) should be greater than a particular limiting value, say, \(\sigma_m\), to prevent dry out. Thus, \(\sigma>\sigma_m+2\rho~RV^2\). Taking \(\sigma_r=\sigma_m+2\rho~RV^2\), The criterion to avoid dry out can be expressed as \(\sigma>\sigma_r\), where \(\sigma_r\) is called as the minimum surface tension required and \(\sigma_m\) denotes a margin. In the present work, along with the solution of \(Q\) and \(R\), \(\sigma\) is varied and the value of \(\sigma\) that just avoids dry out, namely, \(\sigma_{r}\) has been determined. Its value is a function of  \(q(x)\), \(\nu\), \(\beta\) and the properties of the working substance. 
\section{RESULTS AND DISCUSSION}
\subsection*{Range of parameters}
Computations are carried out by varying the heat load (\(q\)) in the range
 \([5-200]~kW/m^2\). The heat transfer coefficient on the outer surface of the condenser has been varied in the range  \([100-7000]~W/m^2K\). The third thermal  parameter varied in this study is the equivalent film heat transfer coefficient \((h_{eq})\), which is varied in the range  \([3000-75000]~W/m^2K\). The geometrical parameter varied is the thickness of the cover plate in the range  \([0.5-15]~mm\).  The angle of elevation of the heat pipe denoted as \(\beta\) is varied in the interval \([0,10]\). The material of the cover plate and the working substance considered are Copper and Water respectively. The  length of  condenser (AE), evaporator (FB) and adiabatic  (EF)  regions are kept constant equal to \(2~cm\), \(2~cm\) and \(6~cm\) respectively. 
\subsection*{Definition of different terms}
The new terms included in the discussion of results of the problem are Heat Spread Factor (HSF), Heat Bypass Factor (HBF), Effective Thermal Conductivity (ETC), heat transfer coefficient required in the outer surface of condenser (\(h_{r,T_{evap}}\)) and the minimum surface tension required \(\sigma_r\). Heat Spread Factor(HSF) is defined as the ratio of heat transferred to liquid from the adiabatic regions spread from evaporator or condenser to total heat. HSF for the evaporator and condenser are computed separately. Bypass factor (HBF) is related to the heat bypassing the liquid and flowing though the solid region of the cover plate. It is defined as  the ratio of the difference between the total heat and the heat transferred to the liquid to the total heat. Effective Thermal Conductivity(ETC) is defined as the ratio of the heat flux to the linear temperature gradient between evaporator and the condenser. The heat transfer coefficient required at the outer surface of the condenser for limiting the average temperature of evaporator to \(T_{evap}\) is denoted as \(h_{r,T_{evap}}\). The minimum surface tension required for the flow of liquid from condenser to evaporator region to avoid dry out condition is denoted as \(\sigma_{r}\). The actual surface tension of the working substance should be greater than \(\sigma_{r}\). 
\subsection*{Effect of  heat transfer coefficient at condenser}
Table \ref{conhc} shows the effect of heat transfer coefficient at the outer surface of condenser (\(h\)) on different parameters at heat loads \(25~kW/m^2\) and \(250~kW/m^2\), corresponding to an equivalent film heat transfer coefficient at the inside surface of \(h_{eq}=15000~ W/(m^2K)\). The temperature at all the different  locations decrease with the increase of \(h\). At the heat load of  \(250~kW/m^2\), a heat transfer coefficient as high as \(7000 W/(m^2K)\) is required at the condenser to limit the temperature of the working fluid to \(350 K\). For  small thickness of cover plate as  \(t=1~mm\),  the HBF is found to be negligible and is found to be unaffected by the heat load. The heat spread (HSF) through the adiabatic regions near the condenser and evaporator are nearly equal and is about \(13\%\)  of the total heat. HSF is found to be independent of \(h\) as well as \(q\). The ETC is found to be insensitive to variations of heat load and \(h\) and is found to be \(2.07\) times that of the thermal conductivity of  Copper. \\
\subsection*{Effect of thickness of the cover plate}
Table \ref{thickvar} shows the effect of the thickness of the  cover plate on heat transfer for \(h=3000 W/m^{2}K\) and \(h_{eq}=15000W/m^{2}K\), for two values of evaporator heat load, viz. \(25 kW/m^2\) and \(250 kW/m^2\). For both the heat loads, an  increase of thickness of the  cover plate from \(0.5~mm\) causes the temperature of the heat pipe to reduce very slightly up to \(10~mm\). For \(t>10~mm\), the temperature of the heat pipe increases by negligible amounts. Variation of temperature of the heat pipe with thickness is negligible,  for the entire range of  thickness of the cover plate studied.  Same trend is observed for the entire range of values of \(h\) also. Further, it can be seen from Table \ref{thickvar},  that increase of thickness tends to spread the heat along the axial direction of the heat pipe, thereby reducing the intensity of heat flux entering the working substance undergoing phase change in the micro-grooves of the heat pipe. Figure \ref{hfxs200} shows the variation of heat flux into the working substance as a function of the axial length of the heat pipe for an evaporator heat load of \(200~kW/m^2\) corresponding to \(h=7000 W/(m^2K)\) and \(h_{eq}=15000~W/(m^2K)\). The reductions in the intensity of heat fluxes at the condenser and evaporator regions are clearly notable.  It is proved later that the spread of heat in the axial direction and the resulting reduction of intensity of heat flux cause a reduction of the  surface tension required \(\sigma_r\). As expected, the value of  \(HBF\) increases with the increase of thickness. Its value is as high as \(24.85\%\) for \(t=15~mm\). 
\subsection*{Effect of liquid film heat transfer coefficient inside the grooves}
Table \ref{filmhc} shows the effect of equivalent heat transfer coefficient of the liquid film inside the grooves  \((h_{eq})\) on different
parameters for heat loads \(25~kW/m^2\) and \(250~kW/m^2\) corresponding to \(h=3000~ W/(m^2K)\). It can be seen that the temperature as well as heat spread  reduces with increase of equivalent liquid film heat transfer coefficient. But the increase of \(h_{eq}\) is not found  to have a significant influence on the temperature of the heat pipe. The maximum percentage reduction in temperature of the heat pipe with the increase of \(h_{eq}\) form \(5000~W/(m^2K)\) to \(50000~W/(m^2K)\) is only  \(14.75\%\) at \(q=250~kW/m^2\).  A comparison between Tables \ref{conhc} and \ref{filmhc} shows that the temperature of the heat pipe is more controlled by the heat transfer coefficient at the condenser (\(h\)) than \(h_{eq}\). But the heat spread is controlled more by \(h_{eq}\) than \(h\). Unlike the effect of thickness, increase of heat spread with a reduction of \(h_{eq}\) is achieved at a penalty of slight increase of temperature. Therefore,  a feasible  solution to increase the spread of heat flux and a consequent reduction in the intensity of heat flux into liquid without the increase of temperature of the heat pipe is to increase the thickness of the cover plate. \\ 
\subsection*{Heat transfer coefficient at condenser}
\begin{figure}[h]
\begin{center}
\begin{center}\includegraphics[scale=0.8]{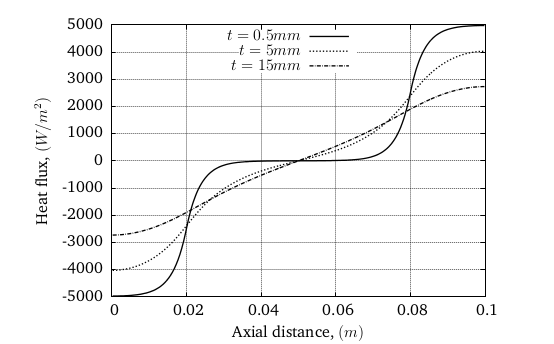}\end{center}
\caption{Variation of heat flux showing the effect of axial spread of heat with changes of thickness corresponding to \(h_{r,350}(t)\) and \(h_{eq}=15000~W/(m^2K)\) for a heat load of \(5~kW/m^2\).}
\label{hfxs5}
\end{center}
\end{figure}

\begin{figure}[h]
\begin{center}
\begin{center}\includegraphics[scale=0.8]{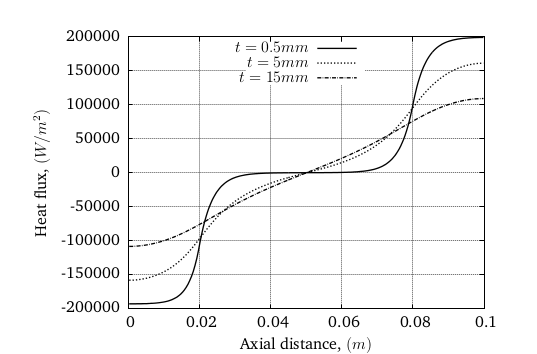}\end{center}
\caption{Variation of heat flux showing the effect of axial spread of heat with changes of thickness corresponding to \(h_{r,350}(t)\) and \(h_{eq}=15000~W/(m^2K)\) for a heat load of \(200~kW/m^2\).}
\label{hfxs200}
\end{center}
\end{figure}

The value of the heat transfer coefficient required (\(h_{r,350}\)) on the outside surface of the heat pipe  to limit the average temperature of the heat source to \(T_{evap}=350K\) is tabulated in Table  (\ref{conhc1}) for different values of heat load (\(q\)) and thickness (\(t\)) of the cover plate, corresponding to \(h_{eq}=15000~W/m^2K\). The value of \(h_{r,350}\) increases with increase of heat load. For a fixed value of \(q\), with increase of \(t\), the value of \(h_{r,350}\) reduces up to  \(t= 5mm\) and then increases. However the value of \(h_{r,350}\) for the thickest cover plate (\(t=15~mm\)) considered in the present study is found to be less than that is required for the  thinnest cover plate (\(t=0.5~mm\)) for all the values of \(q\). Variations of heat flux  into the working substance along the length of the heat pipe corresponding to heat loads \(q=5\) and \(200~kW/m^2\) are given in Figs. (\ref{hfxs5}) and (\ref{hfxs200}) respectively. The three different curves in figures (\ref{hfxs5}) and (\ref{hfxs200}) correspond to  \(t=0.5mm,~5mm\) and \(15mm\) with the convective heat transfer out side the condenser being equal to \(h_{r,350}\) of the respective heat load. It can be seen from the figures that, at low values of \(t\), heat flux is concentrated in the regions of condenser and evaporator with a minimum spread in to the adiabatic region. With increase of \(t\) the  spread of heat in the axial direction increases with a corresponding decrease in the intensity of maximum value of heat flux exchanged to the liquid. Another reason for the reduction in the intensity of the maximum heat flux at higher values of thickness is the by-pass of the heat which amounts to \(24.85\%\) for \(t=15~mm\) as per the conditions in Table \ref{thickvar}. The important parameters related to the heat transfer through the cover plate showing the effect of  thickness of the cover plate and heat loads are tabulated in Table \ref{thermalperf1}.  Clearly, the increase of  heat spread and by-pass with increase of \(t\) from \(0.5~mm\) could be the reasons   for lowering the heat transfer requirement outside the condenser  \(t\). With increase of thickness beyond \(t=5~mm\), the increase of thermal resistance leads to a slight increase of \(h_{r,350}\). The reduction of the maximum value of intensity of heat flux to liquid is expected to reduce the  magnitude of gradients of radius of the liquid meniscus and velocity in the micro-grooved channel.
The effect of the equivalent film  heat transfer coefficient inside the micro grooved channel  on \(h_{r,350}\) has also been investigated  for \(h_{eq}=25000\), \(h_{eq}=50000\) and \(h_{eq}=75000\). The results of this parametric study are presented in Tables \ref{conhc2},\ref{conhc3} and \ref{conhc4} respectively, show that increase of \(h_{eq}\) reduces the heat transfer requirement outside the condenser (\(h_{r,350}\)) for all values of thickness of the cover plate. 
\subsection*{Estimation of equivalent heat transfer coefficient over the inside surface of the microgrooves}
\begin{figure}[h]
\begin{center}
\includegraphics[scale=0.8]{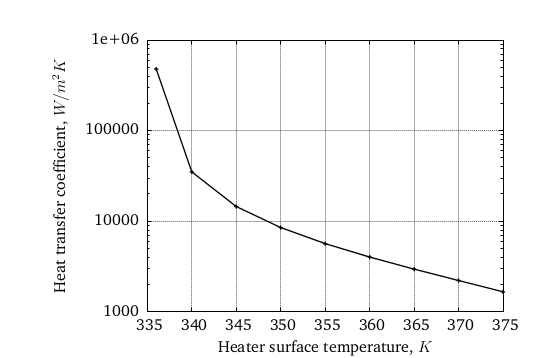}
\caption{Variation of equivalent heat transfer coefficient over the inside surface of the microgrooves with heater surface temperature  corresponding to \(t=5~mm\), \(k=401~W/mK\), \(h=3000~W/m^2K\), \(T_{\infty}=300~K\) and \(q=100~W/m^2\). The \(y\) axis is plotted on \(\log10\) scale.}
\label{heq1pl}
\end{center}
\end{figure}

The equivalent heat transfer coefficient (\(h_{eq}\)) over the inside surface of the microgrooves has been estimated by keeping \(t=5~mm\), \(k=401~W/mK\), \(h=3000~W/m^2K\), \(T_{\infty}=300~K\) and \(q=100~kW/m^2\) and by varying the heater surface temperature in the range \([336-360]~K\). It is found that for the set of fixed conditions  used,  the heater surface temperature \(T_{evap}\) can not be reduced below \(336~K\). Figure (\ref{heq1pl}) shows the variation of \(h_{eq}\) with \(T_{evap}\). The value of \(h_{eq}\) is found to vary widely for small changes of heater surface temperature. Therefore, accuracy of the estimated value depends on the accuracy of the heater surface temperature.
\subsection*{Minimum surface tension required to avoid dry out}
\begin{figure}[h]
\begin{center}
\includegraphics[scale=0.8]{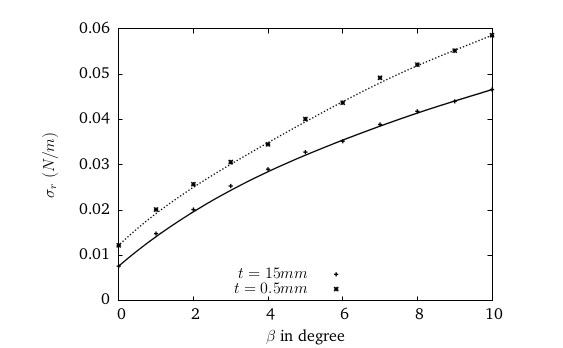}
\caption{Variation of minimum surface tension to avoid the dry out condition with angle of elevation of the heat pipe for two values of thickness of the cover plate,  \(t=0.5mm\) and \(t=15mm\) for \(q=100~kW/m^2\).}
\label{surf100}
\end{center}
\end{figure}

\begin{figure}[h]
\begin{center}
\includegraphics[scale=0.8]{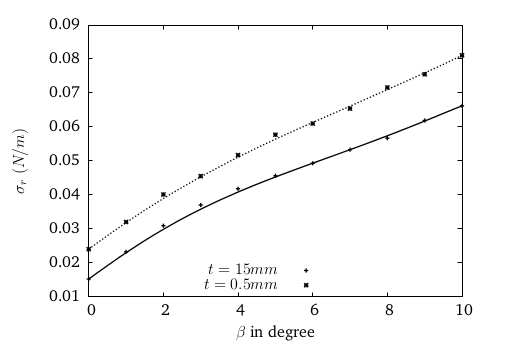}
\caption{Variation of minimum surface tension to avoid the dry out condition with angle of elevation of the heat pipe for two values of thickness of the cover plate,  \(t=0.5mm\) and \(t=15mm\) for \(q=200~kW/m^2\).}
\label{surf200}
\end{center}
\end{figure}

\begin{figure}[h]
\begin{center}
\includegraphics[scale=0.8]{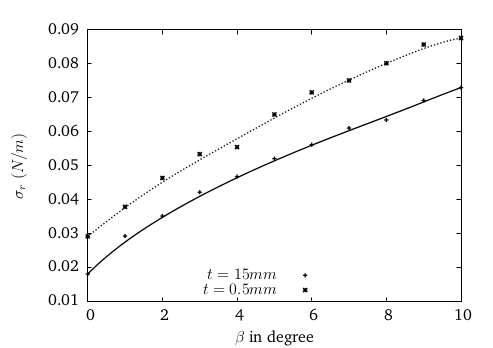}
\caption{Variation of minimum surface tension to avoid the dry out condition with angle of elevation of the heat pipe for two values of thickness of the cover plate,  \(t=0.5mm\) and \(t=15mm\) for \(q=250~kW/m^2\).}
\label{surf250}
\end{center}
\end{figure}

Determination of surface tension required under different operating conditions is an important parameter required for selecting the amount and type of  surfactants to be used while preparing the working substance  of a heat pipe.  In this study water is taken as the base fluid, therefore, the coefficient of viscosity and latent heat of vaporization are taken for water. For constant values of \(\nu\) and \(h_{fg}\), the meniscus radius and velocity of the working fluid depend on  \(\beta\), \(q(x)\) and \(\sigma\).  The effect of the thickness of the cover plate in altering the distribution and maximum intensity of \(q(x)\)  was studied from the analysis of heat transfer through the cover plate. Therefore, the value of minimum surface tension required for avoiding the dry out condition, \(\sigma_r\), has been investigated by varying the thickness of the cover plate, \(t\) and the angle of elevation of the heat pipe, \(\beta\). Figures (\ref{surf100}), (\ref{surf200}) and (\ref{surf250}) show the variation of \(\sigma_r\) with \(\beta\) for \(t=0.5mm\) and \(t=15mm\) corresponding to heat loads of \(100 kW/m^2\),  \(200 kW/m^2\) and \(250 kW/m^2\), respectively. While the value of \(\sigma_r\) increases with increase of \(\beta\), for a constant value of \(\beta\), the value of \(\sigma_r\) is found to reduce with increase of \(t\) in all the cases investigated. A percentage reduction in \(\sigma_r\) of \(20-38\%\) has been found for an increase of \(t\) from \(0.5mm\) to \(15mm\).
\subsection*{Liquid velocity and meniscus radius}
\begin{figure}[h]
\begin{center}
\includegraphics[scale=0.8]{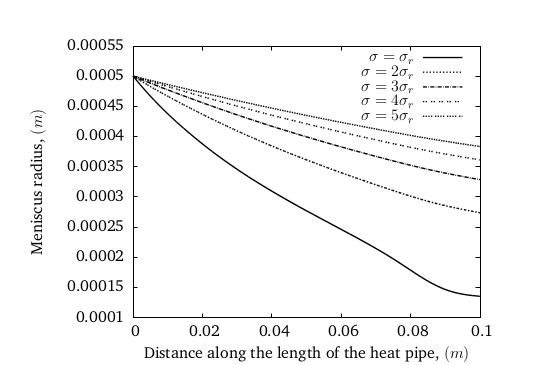}
\caption{Variation of meniscus radius along the axis of the heat pipe for  \(t=0.5mm\), \(q=100~kW/m^2\),  \(\beta=10^{o}\), \(R(x=0)=0.5mm\), \(h=2600~W/m^2K\) and \(h_{eq}=15000~W/m^2K\).}
\label{radtp5}
\end{center}
\end{figure}

\begin{figure}[h]
\begin{center}
\includegraphics[scale=0.8]{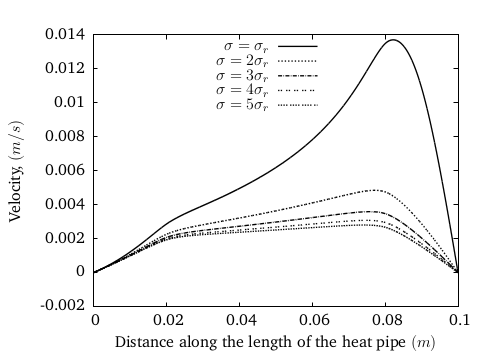}
\caption{Variation of liquid velocity along the axis of the heat pipe for  \(t=0.5mm\), \(q=100~kW/m^2\), \(\beta=10^{o}\), \(h=2600~W/m^2K\) and \(h_{eq}=15000~W/m^2K\).}
\label{veltp5}
\end{center}
\end{figure}

\begin{figure}[h]
\begin{center}
\includegraphics[scale=0.8]{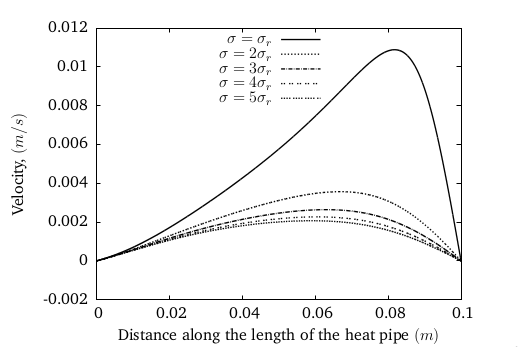}
\caption{Variation of liquid velocity along the axis of the heat pipe for  \(t=15mm\), \(q=100~kW/m^2\),  \(\beta=10^{o}\), \(h=2600~W/m^2K\) and \(h_{eq}=15000~W/m^2K\).}
\label{velt15}
\end{center}
\end{figure}

\begin{figure}[h]
\begin{center}
\includegraphics[scale=0.8]{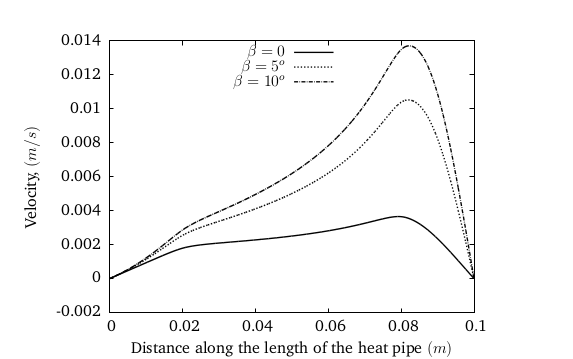}
\caption{Comparison of variation of liquid velocity for different values of \(\beta\) for  \(t=0.5mm\), \(q=100~kW/m^2\), \(\sigma=\sigma_r\), \(h=2600~W/m^2K\) and \(h_{eq}=15000~W/m^2K\).}
\label{betacomp}
\end{center}
\end{figure}

\begin{figure}[h]
\begin{center}
\includegraphics[scale=0.8]{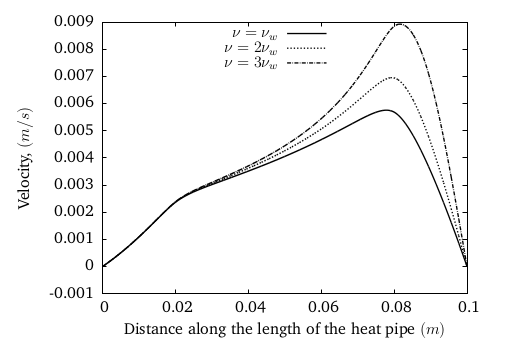}
\caption{Comparison of variation of liquid velocity for different values of viscosity of the working substance for  \(t=0.5mm\), \(q=100~kW/m^2\), \(\sigma=3\sigma_r\), \(h=2600~W/m^2K\) and \(h_{eq}=15000~W/m^2K\).}
\label{visccomp}
\end{center}
\end{figure}

In this section results from the solution of equations of flow of the liquid is presented for a fixed set of thermal parameters by varying the  surface tension of the fluid at \(\beta=10^{o}\).  Figure (\ref{radtp5}) shows the variation of radius of the meniscus of the liquid in the triangular micro groove corresponding to \(q=100~kW/m^2\), \(h=2600~W/m^2K\), \(h_{eq}=15000~W/m^2K\) and \(t=0.5mm\). The surface tension of the working substance is varied as a multiple of \(\sigma_r\). For the lower values of \(\sigma\), the magnitude of denominator of the equation (\ref{thick1}) reduces, therefore, the gradient of \(R\) becomes more negative. This fact is clear by noting the increase of  \(\left|\frac{dR}{dx}(x=0)\right|\) with the decrease of \(\sigma\) in Figure (\ref{radtp5}). For a fixed set of thermal parameters, the distribution of \(q(x)\) remains unchanged. Correspondingly,  with a decrease of radius of liquid meniscus, the velocity of the liquid in the grooves becomes more. This can be observed from the higher  rate of increase of velocity of the liquid in Figure (\ref{veltp5}) for lower values of \(\sigma\). The value of the peak velocity of liquid in the groove increases with decrease  of the surface tension. It is also clear from Figure (\ref{veltp5}) that the magnitude of \(\frac{dV_{max}}{d\sigma}\) is more close to \(\sigma_r\) and the variation of \(V_{max}\) is less pronounced at higher values of \(\sigma\). To observe the effect of variation of the thickness of the cover plate, the distribution of the velocity of liquid has been studied for a thickness of cover plate equal to \(15mm\). The results thus obtained are presented in Figure (\ref{velt15}). The velocity distributions in Figures (\ref{veltp5}) and (\ref{velt15}) are smooth compared to the predictions in the previous investigations \cite{sobhan2000investigations},\cite{suman2005model}. This is due to the comparatively smooth variations of \(q(x)\) computed in the present work. The smoothness in the variation of velocity improves with increase of thickness of the cover plate due to higher axial heat spread and by-pass of heat. \\
Figure (\ref{betacomp}) shows a comparison of variation of liquid velocity for three different angles of elevation of the heat pipe viz. \(\beta=0,5\) and \(10\) degree and \(t=0.5mm\) at \(\sigma=\sigma_r\). It can be observed that the peak velocity of the liquid increases with increase of \(\beta\). The effect of liquid viscosity on the variation of liquid velocity has also been investigated for \(\nu=\nu_{w},2\nu_{w}\) and \(3\nu_{w}\), where \(\nu_{w}\) is the kinematic viscosity of water which is taken as \(0.45\times10^{-6}~m^2/s\). Since \(\sigma_r\) varies with the value of viscosity the effect of viscosity on the liquid velocity has been studied by keeping the surface tension at a constant value equal to \(\sigma_r\) corresponding to \(\nu=3\nu_w\), which is equal to \(0.0985~N/m\). Figure (\ref{visccomp}) shows the variation of liquid velocity with viscosity as the variable parameter. It can be seen that while other conditions remain the same, an increase of viscosity  leads to an increase of velocity of liquid through the channel. 
\section{CONCLUSIONS}
A constant liquid vapor interface temperature of the working substance of a heat pipe is solved by considering the axial spread of heat along the length of the heat pipe and heat bypass through the cover plate. The results show that, the temperature of the heat pipe is  more controlled by the heat transfer coefficient at the condenser and is less influenced by the variations of  equivalent liquid film heat transfer coefficient inside the micro grooves and the thickness of the cover plate. The liquid film heat transfer coefficient and the thickness of the cover plate influence the spread of the heat in the axial direction. Spread of the heat reduces the intensity of heat flux towards the working substance.  \\

Two useful parameters  needed for the design of a heat pipe such as (i) convective heat transfer coefficient required at  the condenser to limit the temperature at the heat source to a specified value and (ii) the minimum surface tension required for the working substance to avoid the dry out condition have been determined under different operating conditions. For a given heat load, the heat transfer coefficient required at the condenser reduces with increase of spread of heat along the length of the heat pipe. The minimum surface tension required for the working substance also reduces with the increase of spread of heat. An increase of thickness of the cover plate  has  positive effects of reducing the heat transfer and surface tension requirements. The study demonstrated a methodology for the estimation of the equivalent heat transfer coefficient (\(h_{eq}\)) of the liquid film inside the micro groove without involving any direct measurement of variables inside the microgrooves. As  the variables outside the heat pipe can be accurately measured, the present methodology consisting of a heat conduction analysis provides a means for an accurate determination of the average heat transfer coefficient in boiling and condensation of the working substance.\\
 
The spread of heat along the axial direction of the heat pipe has an effect of smoothening the variation of velocity of  flow of the working substance. An increase of the surface tension  reduces the velocity of the working substance. For a given value of surface tension of the liquid, an increase of the coefficient of viscosity leads to a reduction of the velocity of flow.  
\section*{REFERENCES}
\bibliographystyle{asmems4}
\bibliography{reference.bib}

\clearpage

\begin{table}[h]
\caption{Effect of heat transfer coefficient at the condenser on the thermal performance of the heat pipe corresponding to heat fluxes \(q=25kW/m^2\) and \(q=250 kW/m^2\). (Other parameters: thickness of the cover plate \(t=1mm\), liquid film side heat transfer coefficient inside the heat pipe \(h_{eq}=15000~W/m^{2}K\))}
\begin{center}
\resizebox{\columnwidth}{!}{%
 \begin{tabular}{cccccccc}
 \toprule
 &\multicolumn{3}{c}{Temperature}&&&\multicolumn{2}{c}{HSF} \\
 \cmidrule{2-4}\cmidrule{7-8} 
\(h\)& Condenser &Evaporator & Interface & \(K_{eff}\) & HBF & Evaporator & Condenser  \\
\hline
\(\frac{W}{(m^{2}K)}\)&K&K&K&\(\frac{W}{(mK)}\)&\%&\%&\%\\
\bottomrule
\multicolumn{8}{c}{\(q=25 kW/m^2\)} \\
\hline
500 & 349.98 & 353.00 & 351.50 & 826.43 & 0.0821 & 12.99 & 13.04\\
1000 & 324.98 & 328.00 & 326.50 & 826.70 & 0.0823 & 12.99 & 13.09 \\
2000 & 312.48 & 315.50 & 314.00 & 827.23 & 0.0826 & 12.99 & 13.19 \\
3000 & 308.31 &  311.33 & 309.83 & 827.74 & 0.0829 & 12.99 & 13.29 \\
4000 & 306.23 & 309.25 & 307.75 & 828.22 & 0.0832 & 12.99 & 13.38 \\
5000 & 304.98 & 308.00 & 306.49 & 828.68 & 0.0834 & 12.99 & 13.46 \\
6000 & 304.15 & 307.16 & 305.66 & 829.11 & 0.0837 & 12.99 & 13.55 \\
7000 & 303.55 & 306.57 & 305.06 & 829.53 & 0.0839 & 12.99 & 13.63 \\
\bottomrule
\multicolumn{8}{c}{\(q=250 kW/m^2\)} \\
\hline
500 & 799.84 & 830.09 & 815.06 & 826.43 & 0.0821 & 12.99 & 13.04\\
2000 & 424.85 & 455.07 & 440.03 & 827.23 & 0.0826 & 12.99 & 13.19 \\
3000 & 383.18 &  413.39 & 398.35 & 827.74 & 0.0829 & 12.99 & 13.29 \\
4000 & 362.35 & 392.54 & 377.50 & 828.22 & 0.0832 & 12.99 & 13.38 \\
5000 & 349.85 & 380.02 & 364.98 & 828.68 & 0.0834 & 12.99 & 13.46 \\
6000 & 341.64 & 371.67 & 356.64 & 829.11 & 0.0837 & 12.99 & 13.55 \\
7000 & 335.57 & 365.71 & 350.67 & 829.53 & 0.0839 & 12.99 & 13.63 \\
\bottomrule
\end {tabular}
}
\end{center}
\label{conhc}
\end{table} 

\begin{table}[h]
\caption{Effect of thickness of the copper cover plate on the thermal performance of the heat pipe corresponding to heat fluxes \(q=25kW/m^2\) and \(q=250~kW/m^2\). (Other parameters: heat transfer coefficient at condenser \(h=3000~W/(m^2K)\), liquid film side heat transfer coefficient inside the heat pipe \(h_{eq}=15000~W/m^{2}K\))}
\begin{center}
\resizebox{\columnwidth}{!}{%
 \begin{tabular}{cccccccc}
 \toprule
 &\multicolumn{3}{c}{Temperature}&&&\multicolumn{2}{c}{HSF} \\
 \cmidrule{2-4}\cmidrule{7-8} 
\(t\) & Condenser &Evaporator & Interface & \(K_{eff}\) & HBF & Evaporator & Condenser  \\
\hline
\((mm)\)&K&K&K&\(\frac{W}{(mK)}\)&\%&\%&\%\\
\bottomrule
\multicolumn{8}{c}{\(q=25 kW/m^2\)} \\
\hline
0.5 & 308.32 & 311.41 & 309.87 & 807.72 & 0.0052 & 9.17 & 9.44\\
1 & 308.32 & 311.33 & 309.83 & 827.74 & 0.0829 & 12.99 & 13.29 \\
1.5 & 308.32 & 311.29 & 309.81 & 840.67 & 0.2974 & 14.80 &16.12 \\
2 & 308.32 & 311.26 & 309.79 & 849.88 & 0.6534 & 18.09 & 18.38 \\
5 &308.32&311.19&309.76&871.64&4.79&26.11&26.32 \\
10 &308.32&311.20&309.76&868.43&14.61&31.96&32.10 \\
15 &308.32&311.24&309.78&856.42&24.85&34.95&35.06 \\
\bottomrule
\multicolumn{8}{c}{\(q=250 kW/m^2\)} \\
\hline
0.5 & 383.17 & 414.13 & 398.72 & 807.72 & 0.0052 & 9.17 & 9.44\\
1 & 383.18 & 413.39 & 398.35 & 827.74 & 0.0829 & 12.99 & 13.29 \\
1.5 & 383.19 & 412.93 & 398.12 & 840.67 & 0.2974 & 14.80 &16.12 \\
2 & 383.20 & 412.61 & 397.96 & 849.88 & 0.6534 & 18.09 & 18.38 \\
5 &383.22&411.90&397.61&871.64&4.79&26.11&26.32 \\
10 &383.23&412.02&397.67&868.43&14.61&31.96&32.10 \\
15 &383.23&412.42&397.88&856.42&24.85&34.95&35.06 \\
\bottomrule
\end {tabular}
}
\end{center}
\label{thickvar}
\end{table}

\begin{table}[h]
\caption{Effect of  liquid film side heat transfer coefficient on the thermal performance of the heat pipe corresponding to heat fluxes \(q=25kW/m^2\) and \(q=250 kW/m^2\). (Other parameters: thickness of the cover plate \(t=1mm\), heat transfer coefficient at condenser \(h=3000 W/m^2K\).)}
\begin{center}
\resizebox{\columnwidth}{!}{%
 \begin{tabular}{cccccccc}
 \toprule
 &\multicolumn{3}{c}{Temperature}&&&\multicolumn{2}{c}{HSF} \\
 \cmidrule{2-4}\cmidrule{7-8} 
\(h_{eq}\)& Condenser &Evaporator & Interface & \(K_{eff}\) & HBF & Evaporator & Condenser  \\
\hline
\(\frac{W}{(m^{2}K)}\)&K&K&K&\(\frac{W}{(mK)}\)&\%&\%&\%\\
\bottomrule
\multicolumn{8}{c}{\(q=25 kW/m^2\)} \\
\hline
5000 & 308.30 & 316.17 & 312.24 & 317.45 & 1.593 & 20.98 & 21.66\\
10000 & 308.31 & 312.63 & 310.48 & 578.39 & 0.2885 & 15.68 & 16.10 \\
20000 & 308.32 & 310.66 & 309.49 & 1068.53 & 0.0299 & 11.33 & 11.55 \\
30000 & 308.32 &  309.96 & 309.14 & 1528.96 & 0.0057 & 9.34 & 9.48 \\
40000 & 308.32 & 309.59 & 308.96 & 1965.04 & 0.0014 & 8.14 & 8.24 \\
50000 & 308.32 & 309.37 & 308.85 & 2379.63 & 0.0004 & 7.33 & 7.41 \\
\bottomrule
\multicolumn{8}{c}{\(q=250 kW/m^2\)} \\
\hline
5000 & 383.01 & 461.76 & 422.42 & 317.45 & 1.59 & 20.98 & 21.66\\
10000 & 383.13 & 426.35 & 404.81 & 578.39 & 0.2885 & 15.68 & 16.10 \\
20000 & 383.21 & 406.61 & 394.97 & 1068.53 & 0.0299 & 11.33 & 11.55 \\
30000 & 383.25 &  399.60 & 391.46  & 1528.96 & 0.0057 & 9.34 & 9.48 \\
40000 & 383.26 & 395.99 & 389.66 & 1965.04 & 0.0014 & 8.14 & 8.24 \\
50000 & 383.28 & 393.78 & 388.56& 2379.63 & 0.0004 & 7.33 & 7.41 \\
\bottomrule
\end {tabular}
}
\end{center}
\label{filmhc}
\end{table}
\begin{table}
\caption{The computed heat transfer coefficient at the outer surface of the condenser to limit the temperature of the heater surface to \(350K\), for different values of  heat load (\(q\)) and thickness of the cover plate. The liquid film side heat transfer coefficient inside the heat pipe is, \(h_{eq}=15000~W/m^{2}K\)}
\vspace{0.3cm}
\begin{center}
\resizebox{\columnwidth}{!}{%
 \begin{tabular}{cccccccc}
 \toprule
\(q\) &\multicolumn{7}{c}{Heat transfer coefficient outside the condenser, \(\left(h~\frac{W}{m^2K}\right)\)  } \\
\hline
\((kW/m^2)\)& \(t=0.5~mm\) &\(t=1~mm\) &\(t=1.5~mm\) & \(t=2~mm\) & \(t=5~mm\)& \(t=10~mm\) & \(t=15~mm\)  \\
\bottomrule
5 & 101.28 & 101.21 & 101.17 & 101.20 & 101.15 & 101.15 & 101.18\\
10 & 204.99 & 204.92 & 204.74 & 204.74& 204.67 & 204.70 & 204.74 \\
15 & 311.46 & 311.18 & 311.04 & 310.92 & 310.62 & 310.75 & 310.83 \\
20 & 420.70 &  420.14 & 419.87 & 419.66 & 419.14 & 419.22 & 419.52 \\
25 & 532.83 & 532.05 & 531.57 & 531.24 & 530.31 & 530.36 & 530.95 \\
50 & 1140.69 & 1136.57 & 1134.29 & 1132.76 & 1129.07 & 1129.71 & 1131.72 \\
100 & 2655.99 & 2632.16 & 2618.07 & 2610.82 & 2592.33 & 2593.25 & 2605.90 \\
200 & 7879.70 & 7686.17 & 7578.84 & 7501.03 & 7350.14 & 7375.77 & 7465.18 \\
\bottomrule
\end {tabular}
}
\end{center}
\label{conhc1}
\end{table}

\begin{table}[h]
\caption{Effect of thickness on the thermal performance of heat pipe corresponding to \(h_{r,350}\) for \(q=25kW/m^2\) and \(h_{eq}=15000 W/(m^2K)\)}
\begin{center}
\resizebox{\columnwidth}{!}{%
 \begin{tabular}{ccccccccc}
 \toprule
 &\multicolumn{3}{c}{Temperature}&&&\multicolumn{2}{c}{HSF} &\\
 \cmidrule{2-4}\cmidrule{7-8} 
\(t\)& Condenser &Evaporator & Interface & \(K_{eff}\) & HBF & Evaporator & Condenser&\(h_{r,350}\)  \\
\hline
\(mm\)&K&K&K&\(\frac{W}{(mK)}\)&\%&\%&\%& \(\frac{W}{(m^{2}K)}\)\\
\bottomrule
\multicolumn{9}{c}{\(q=5 kW/m^2\)} \\
\hline
0.5 & 349.36 & 349.98 & 349.67 & 806.39 & 0.0051& 9.17 & 9.18&101.28\\
5 &349.42&350.00&349.71&870.38&4.778&26.12&26.13&101.15 \\
15 &349.41&349.99&349.70&855.55&24.82&34.97&34.97&101.18 \\
\bottomrule
\multicolumn{9}{c}{\(q=25kW/m^2\)} \\
\hline
0.5 & 346.90 & 350.00 & 348.46 & 806.60 & 0.0051& 9.17 & 9.22&532.83\\
5 &347.13&350.02&348.57&870.58&4.781&26.12&26.16&530.31 \\
15 &347.07&349.99&348.54&855.69&24.83&34.96&34.98&530.95 \\
\bottomrule
\multicolumn{9}{c}{\(q=250kW/m^2\)} \\
\hline
0.5 & 325.26 & 349.97 & 337.64 & 809.57 & 0.0053 & 9.17 & 9.79&7879.70\\
5 &327.12&350.02&338.59&873.34&4.816&26.10&26.58&7350.14 \\
15 &326.71&350.03&338.40&857.65&24.88&34.93&35.18&7465.18 \\
\bottomrule

\end {tabular}
}
\end{center}
\label{thermalperf1}
\end{table}

\begin{table}[h]
\caption{The computed heat transfer coefficient at the outer surface of the condenser to limit the temperature of the heater surface to \(350K\), for different values of  heat load (\(q\)) and thickness of the cover plate. The liquid film side heat transfer coefficient inside the heat pipe is, \(h_{eq}=25000~W/m^{2}K\)}
\begin{center}
\resizebox{\columnwidth}{!}{%
 \begin{tabular}{cccccccc}
 \toprule
\(q\) &\multicolumn{7}{c}{Heat transfer coefficient outside the condenser, \(\left(h~\frac{W}{m^2K}\right)\)  } \\
\hline
\((kW/m^2)\)& \(t=0.5~mm\) &\(t=1~mm\) &\(t=1.5~mm\) & \(t=2~mm\) & \(t=5~mm\)& \(t=10~mm\) & \(t=15~mm\)  \\
\bottomrule
5 & 100.76 & 100.76 & 100.76 & 100.78 & 100.84 & 100.91 & 100.98\\
10 & 203.11 & 203.08 & 203.11 & 203.13& 203.32 & 203.68 & 205.04 \\
15 & 307.07 & 307.03 & 307.07 & 307.12 & 307.56 & 308.33 & 309.08 \\
20 & 412.65 &  412.60 & 412.66 & 412.71 & 413.56 & 414.99 & 416.24 \\
25 & 519.92 & 519.87 & 519.85 & 519.87 & 521.34 & 523.67 & 525.78 \\
50 & 1083.08 & 1082.75 & 1083.07 & 1083.48 & 1089.19 & 1099.40 & 1108.65 \\
100 & 2361.96 & 2360.70 & 2361.72 & 2365.02 & 2391.58 & 2441.38 & 2487.38 \\
200 & 5765.78 & 5757.68 & 5766.53 & 5783.58 & 5944.63 & 6264.1 & 6571.7 \\
\bottomrule
\end {tabular}
}
\end{center}
\label{conhc2}
\end{table}
\begin{table}[h]
\caption{The computed heat transfer coefficient at the outer surface of the condenser to limit the temperature of the heater surface to \(350K\), for different values of  heat load (\(q\)) and thickness of the cover plate. The liquid film side heat transfer coefficient inside the heat pipe is, \(h_{eq}=50000~W/m^{2}K\)}
\vspace{0.3cm}
\begin{center}
\resizebox{\columnwidth}{!}{%
 \begin{tabular}{cccccccc}
 \toprule
\(q\) &\multicolumn{7}{c}{Heat transfer coefficient outside the condenser, \(\left(h~\frac{W}{m^2K}\right)\)  } \\
\hline
\((kW/m^2)\)& \(t=0.5~mm\) &\(t=1~mm\) &\(t=1.5~mm\) & \(t=2~mm\) & \(t=5~mm\)& \(t=10~mm\) & \(t=15~mm\)  \\
\bottomrule
5 & 100.39 & 100.43 & 100.43 & 100.45 & 100.54 & 100.68 & 100.81\\
10 & 201.60 & 201.68 & 201.73 & 201.78& 202.20 & 202.80 & 203.28 \\
15 & 303.75 & 303.83 & 303.93 & 304.13 & 304.98 & 306.33 & 307.46 \\
20 & 406.65 &  406.87 & 407.05 & 407.33 & 408.93 & 411.36 & 413.44 \\
25 & 510.28 & 510.68 & 511.09 & 511.53 & 513.97 & 517.94 & 521.25 \\
50 & 1041.56 & 1043.19 & 1045.57 & 1047.13 & 1057.93 & 1074.01 & 1088.07 \\
100 & 2177.03 & 2182.64 & 2189.88 & 2198.15 & 2245.02 & 2319.91 & 2385.17 \\
200 & 4770.65 & 4804.76 & 4839.18 & 4877.25 & 5119.63 & 5522.01 & 5913.80 \\
\bottomrule
\end {tabular}
}
\end{center}
\label{conhc3}
\end{table}

\begin{table}[t]
\caption{The computed heat transfer coefficient at the outer surface of the condenser to limit the temperature of the heater surface to \(350K\), for different values of  heat load (\(q\)) and thickness of the cover plate. The liquid film side heat transfer coefficient inside the heat pipe is, \(h_{eq}=75000~W/m^{2}K\)}
\vspace{0.3cm}
\vspace{0.3cm}
\begin{center}
\resizebox{\columnwidth}{!}{%
 \begin{tabular}{cccccccc}
 \toprule
\(q\) &\multicolumn{7}{c}{Heat transfer coefficient outside the condenser, \(\left(h~\frac{W}{m^2K}\right)\)  } \\
\hline
\((kW/m^2)\)& \(t=0.5~mm\) &\(t=1~mm\) &\(t=1.5~mm\) & \(t=2~mm\) & \(t=5~mm\)& \(t=10~mm\) & \(t=15~mm\)  \\
\bottomrule
5 & 100.26 & 100.26 & 100.31 & 100.33 & 100.44 & 100.61 & 100.74\\
10 & 201.12 & 201.18 & 201.24 & 201.38& 201.80 & 202.46 & 202.97 \\
15 & 302.54 & 302.71 & 302.89 & 303.05 & 304.04 & 305.56 & 306.91 \\
20 & 404.53 &  404.84 & 405.09 & 405.47 & 407.28 & 409.95 & 412.28 \\
25 & 507.11 & 507.59 & 508.06 & 508.58 & 511.46 & 515.60 & 519.42 \\
50 & 1028.99 & 1030.75 & 1032.61 & 1034.74 & 1046.86 & 1064.84 & 1080.39 \\
100 & 2119.20 & 2127.51 & 2135.69 & 2144.67 & 2196.43 & 2278.07 & 2349.58 \\
200 & 4504.24 & 4542.50 & 4582.11 & 4623.20 & 4868.39 & 5287.33 & 5692.50 \\
\bottomrule
\end {tabular}
}
\end{center}
\label{conhc4}
\end{table}

\end{document}